\documentclass[11pt]{article}
\usepackage{BSMvietnam,epsfig}

\def\be{\begin{equation}}
\def\ee{\end{equation}}
\def\bea{\begin{eqnarray}}
\def\eea{\end{eqnarray}}

\begin{document}

\vspace*{3.5cm}
\title{TOP QUARK POLARIZATION AND THE SEARCH FOR NEW PHYSICS \footnote{Argonne preprint ANL-HEP-CP-12-87.  Invited paper presented at Beyond the Standard Model of Particle Physics,  Quy Nhon, Vietnam, July 15 -- 21, 2012.  To appear in the Proceedings.}}

\author{ Edmond L. Berger}

\address{High Energy Physics Division, Argonne National Laboratory, Argonne, IL 60439, USA}

\maketitle\abstracts{
Forward-backward asymmetries $A_{FB}^t$ and $A_{FB}^\ell$ are observed in the top quark $t$ rapidity 
distribution and in the rapidity distribution of charged leptons $\ell$ from top quark decay at the Tevatron 
proton-antiproton collider, and a charge asymmetry $A_C$ is seen in proton-proton collisions at the Large 
Hadron Collider (LHC).   In this presentation, I summarize research my collaborators and I have done on the 
interpretation and implications of the Tevatron asymmetries and provide expectations for $A_C$ at the LHC.   
The two asymmetries $A_{FB}^t$ and $A_{FB}^\ell$ are connected through the $(V-A)$ spin correlation 
between the charged lepton and the top quark with different polarization states.  The ratio of the two 
asymmetries provides independent insight into the physics interpretation of the top quark asymmetry.   
A new physics model which produces more right-handed than left-handed top quarks appears to be 
indicated by the present Tevatron data, but an improvement in precision is desirable.}

\section{Introduction}
The large mass of the top quark, of order the electroweak scale, suggests that the top quark may be sensitive 
to electroweak symmetry breaking and to physics beyond-the-standard model.  Experimentally, the observation 
of a larger than expected forward-backward asymmetry $A_{FB}^t$  in the rapidity of top quarks produced at the 
Fermilab Tevatron collider~\cite{Aaltonen:2011kc,Abazov:2011rq} continues to hold considerable attention.   
The asymmetry is defined as
\be
A_{FB}^t = \frac{N(\Delta y>0) - N(\Delta y < 0)}{N(\Delta y>0) + N(\Delta y < 0)},
\label{eq:def_afbt}
\ee
where $\Delta y=y_t-y_{\bar t}$ is the difference between the rapidities of the top quark and the anti-top quark, 
and $N(\Delta y>0)$ ($N(\Delta y < 0 )$) is the number of events with $\Delta y>0$ ($\Delta y<0$).  The proton 
beam is chosen as the direction of positive $z$.  In the standard model (SM), the asymmetry is induced by 
perturbative quantum chromodynamics (QCD) processes beyond the leading order.  

The enhanced asymmetry is one of few manifestations of a deviation from predictions of the SM.     
Many models of new physics (NP) have been proposed to explain the data.  A lengthly list of references and a 
discussion of constraints on these models may be found in Berger {\em et al.}~\cite{Berger:2012tj} and 
in Cvetic {\em et al.}~\cite{Cvetic:2012kv}.  Some of the NP models postulate the existence of new states 
with right-handed couplings of the top quark.  

The large mass of the top quark is important in another respect.  Its short lifetime means that the top quark decays 
as a ``bare" quark.  Its polarization information is  retained in the weak decay $t \rightarrow b \ell \nu$, passed 
to its decay products.   The lepton $\ell$ angular distribution in the top quark rest frame is maximally correlated with 
the top quark spin orientation, providing the opportunity to test non-standard features of NP models such as 
right-handed couplings.   Another method to measure the polarization is based on the lepton momentum 
distribution~\cite{Berger:2012an} and is valuable for use with complex final states in which the top quark rest frame 
is hard to determine.  Of particular interest to us have been the implications of models of new physics for the polarization 
of the top quark, and methods that can be used to measure the polarization~\cite{Berger:2012an}.  In the SM, strong 
production of $t \bar{t}$ pairs in quantum chromodynamics (QCD) yields an equal number of positive and negative 
helicity top quarks, hereafter referred to as $t_R$ and $t_L$.  Electroweak production in single top quark production, 
for example, yields primarily $t_L$.  Therefore, a demonstration that a significant fraction of top quarks are produced 
with positive helicity would herald new physics.   

In addition to $A_{FB}^t$ of the top quark, the D0 group reports a positive forward-backward asymmetry of charged 
leptons from top quark decays.   The measurement is done in two ways~\cite{Abazov:2011rq,:2012bfa}, both based 
on data corresponding to an integrated luminosity of $5.4\rm{~fb}^{-1}$.   The value $A_{FB}^\ell=(15.2\pm 4.0)\%$ 
is measured in the {$\ell$+jets} final states~\cite{Abazov:2011rq}.  The second method uses the dilepton final states 
from $t \bar{t}$ production, where the $W$ bosons from the $t$ and $\bar{t}$ decays both decay leptonically, and the 
result obtained is $A^{\ell}_{\rm FB}= (5.8 \pm 5.1 ({\rm stat}) \pm 1.3 ({\rm syst}))$\%.  A combination of the two 
measurements yields $A^{\ell}_{\rm FB}= {11.8 \pm 3.2}\%$.   The combined result may be compared with the values 
$(2.1\pm0.1)\%$ from simulations of the SM or $(4.7 \pm 0.1)\%$ once QCD+EW corrections 
are included~\cite{Bernreuther:2010ny,:2012bfa}, an excess at the level of 2.2 standard deviations.   The fact that 
$A_{FB}^\ell$ and $ A_{FB}^t$ are larger than the SM predictions indicates that the charged lepton strongly prefers 
to move in the same direction as the top quark from which it originates.   Data on the ratio of the two asymmetries 
tend to favor models in which more $t_R$ than $t_L$ are produced~\cite{Berger:2012tj}, but confirmation with greater 
statistical and systematic precision is desirable.  

In Sec.~\ref{sec:tevatron} the asymmetries measured at the Tevatron are defined and our fits in the framework of $Z'$, 
$W'$, and axigluon new physics models are discussed.    The LHC proton-proton collider offers no preferred 
direction for the measurement a rapidity asymmetry.   Nevertheless, charge asymmetries $A_C^t$ for top quarks 
and $A_C^\ell$ for leptons can be defined and computed.  Using data from the Tevatron, we may obtain expectations 
for these charge asymmetries, and we compare these expectations with LHC data in Sec.~\ref{sec:lhc}.   
Despite limited statistics, the LHC data on the charge asymmetry are also consistent with a deviation from the SM, 
although perhaps not as great a deviation as expected from an extrapolation of the Tevatron observations.      

The relationship of $A_{FB}^t$ and $A_{FB}^\ell$ is addressed in Sec.~\ref{sec:landtasy}.  The essential starting 
point is the $V-A$ structure of the matrix element for the decay $t \rightarrow W^+ b \rightarrow b \ell^+ \nu$.   
We pay particular attention to the positive/negative helicity state of the top quark because the final momentum and angular 
distributions of leptons in the laboratory frame depend significantly on the top quark's polarization state.  We derive 
the relationship of the lepton asymmetry $A_{FB}^\ell$ and the top quark asymmetry $A_{FB}^t$ separately for the 
left- and right-handed polarization states of the top quark.   Different models of new physics produce top quarks with 
different proportions of left- and right-handed polarization.  
For example, $W'$ models produce predominantly right-handed top quarks, whereas the axigluon model 
generates unpolarized top quarks.  We use an axigluon model and a $W^{\prime}$ model in Sec.~\ref{sec:landtasy} 
to illustrate their different expectations for the ratio of the lepton and top quark asymmetries.  

\section{Tevatron Data and Interpretations}
\label{sec:tevatron}

In Berger {\em et al.}~\cite{Berger:2012tj}, we present fits for three models: flavor-changing $Z'$ 
exchange, flavor-changing $W'$ exchange, and axigluon models.   The minimal version of the $Z'$ model 
implies a large rate for same-sign top quark pair production at the LHC, not supported by 
data~\cite{Berger:2011ua,Aad:2012bb,Chatrchyan:2012sa}.  The $W'$ model is constrained by data on 
the $t\bar t$ plus jets final state at the LHC~\cite{Knapen:2011hu,Duffty:2012zz,Chatrchyan:2012su}.
The absence of pronounced deviations from the SM expectation in the measured
$m_{t\bar{t}}$ invariant mass distribution indicates to us that the axigluon should be heavy 
and/or broad.   Another possibility would be to place it below $m_{t\bar{t}}$ threshold~\cite{Gross:2012bz}.

We fit data at the Tevatron to determine the parameters of the three new physics models.  
under consideration.  We scan the parameter space of the models requiring that the predictions fit the 
total cross section as well as CDF data on $A_{FB}^t$  for both $m_{t \bar t} <~450~{\rm GeV}$ and 
$m_{t\bar t} \geq~450~{\rm GeV}$ within $2\sigma$ accuracy.  
\begin{figure}[!htb]
\includegraphics[scale=0.35]{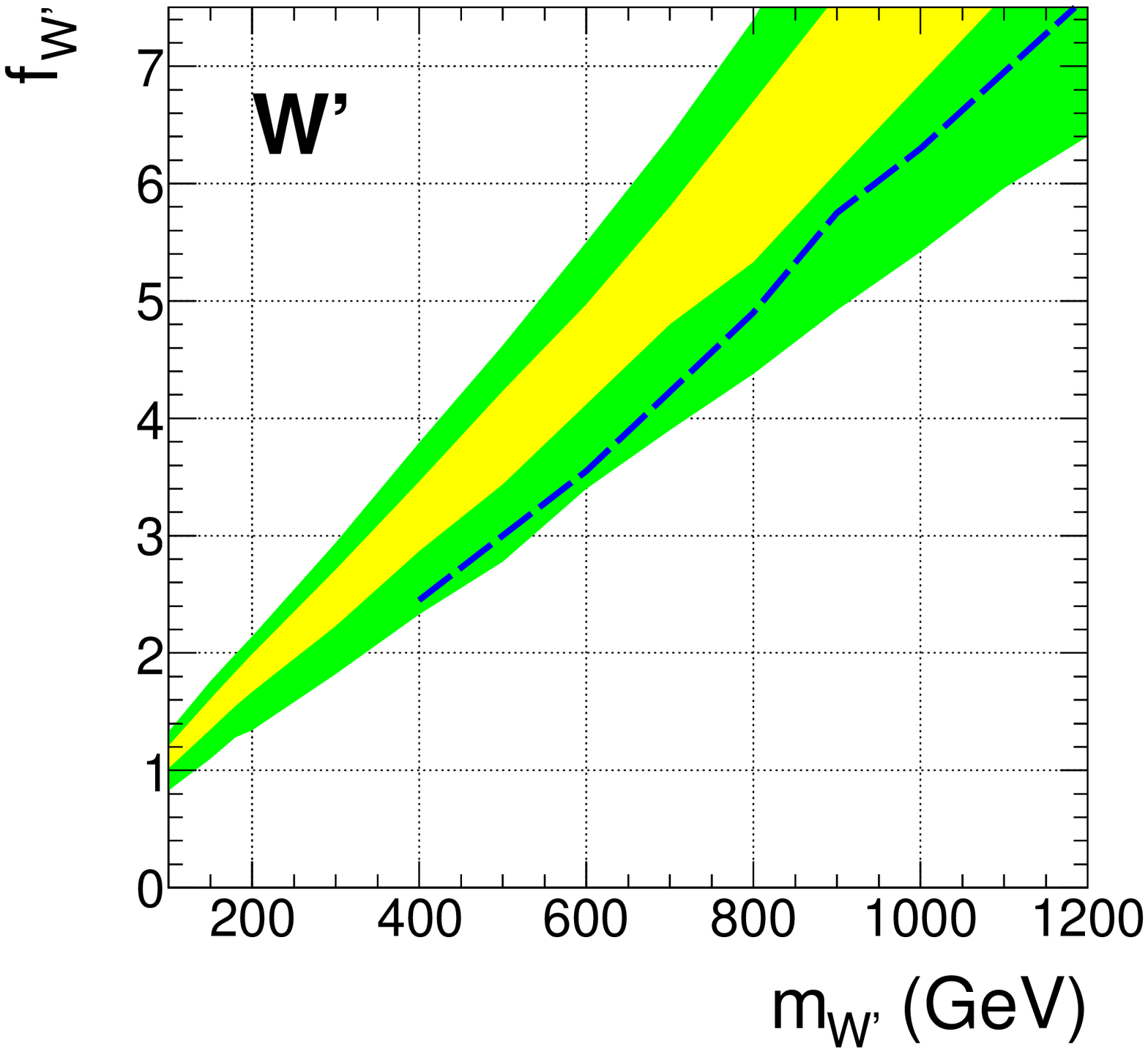}
\includegraphics[scale=0.35]{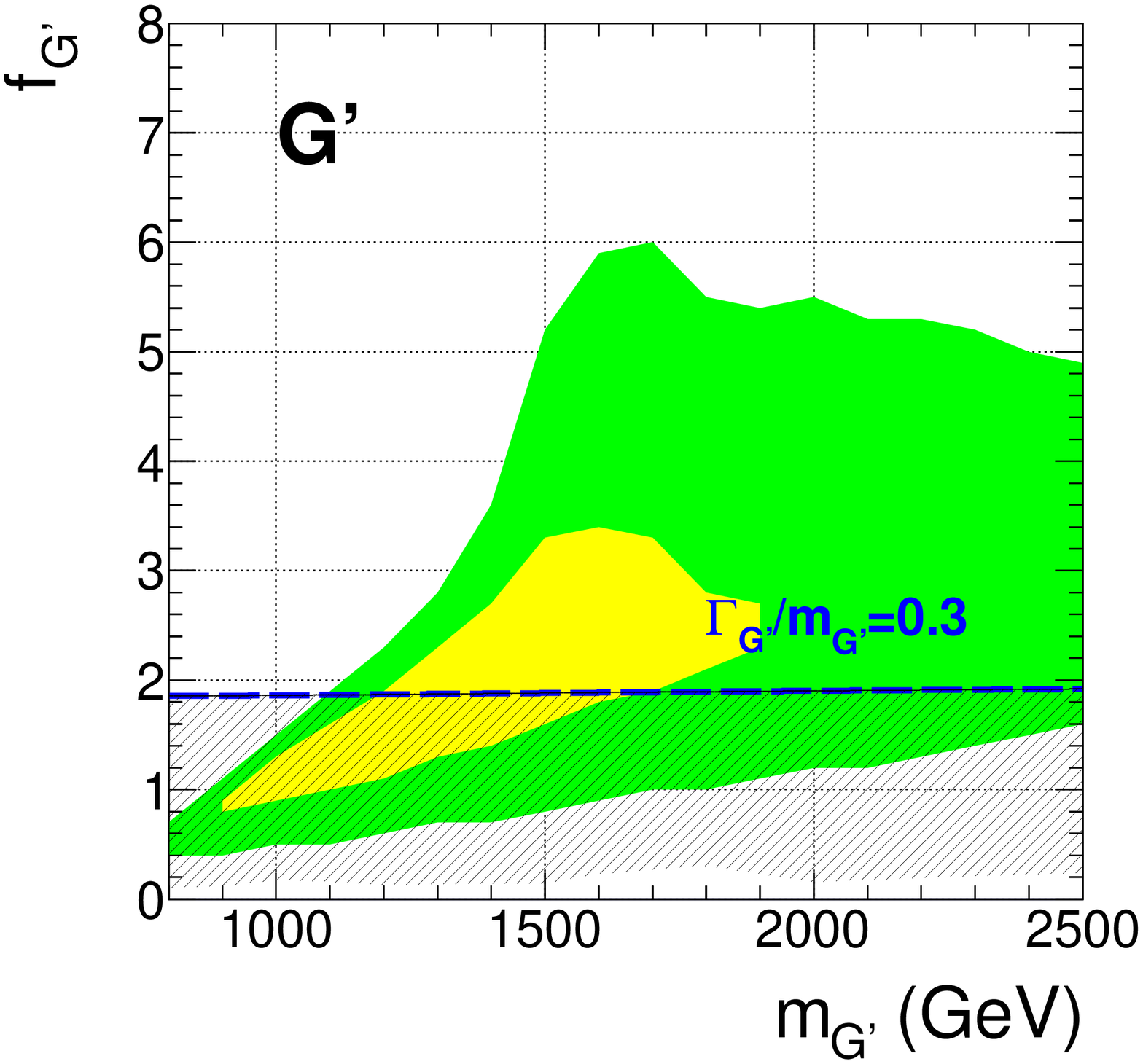}
\caption{The parameter space of two new physics models determined from fits to 
the Tevatron $t\bar t$ total cross section and $A_{FB}^t$ measured by the CDF 
collaboration in the intervals $m_{t\bar t} <~450~{\rm GeV}$ and $m_{t\bar t} 
\geq~450~{\rm GeV}$.  The yellow region fits the data within $1\sigma$ and the green 
region fits within $2\sigma$: flavor-changing 
$W^\prime$ model, and axigluon model. The dashed 
line in the $W^\prime$ case shows the bound on the coupling $f_{W'}$ obtained from an 
analysis of the CMS data on top-pair-plus-one-jet events at the LHC.
The blue shaded region in the axigluon case is inferred from the limits set by ATLAS on axigluons from the search 
for enhancements in the dijet mass distribution.}
\label{fig:tevfit}
\end{figure}

Figure~\ref{fig:tevfit} shows the results of our fits for two of the models.   The fit for the 
$Z'$ model may be found in Berger {\em et al.}~\cite{Berger:2012tj}.  There is a large region of 
parameter space in which 
the $W'$ model can fit the Tevatron data  within $1\sigma$ and $2\sigma$.  However, the region 
above the blue dashed curve is not allowed since too many $t\bar t+j$ events would be produced.  
In the axigluon case, to achieve good agreement with data at the $1\sigma$ level, the mass of 
axigluon is required to be in the range of about $900~{\rm GeV}$ to $1900~{\rm GeV}$.  
For other axigluon masses, the model can only fit data at the $2\sigma$ level.  
Also shown are some bounds on axigluon masses and couplings obtained from a 
search for resonances in the dijet invariant mass distribution~\cite{Aad:2011aj,Berger:2012tj}.
 
\section{LHC Data and Expectations}
\label{sec:lhc}

The LHC proton-proton collider is symmetric in rapidity, and it is ambiguous to define a forward or backward 
region.  However, the $u$ and  $d$ parton densities carry, on average, a larger fraction of the momentum of 
the proton than the $u$ and $d$ antiquark densities.  With the knowledge that there is a forward-backward 
asymmetry in the perturbative production process for $\bar{q} q \rightarrow t\bar t$ production, we 
expect that the top quark at the LHC will be boosted in the direction of the incident quark.  As a result, top 
quarks should accumulate in the region of large rapidity and anti-top quarks will be preferentially in the central 
region.   An asymmetry $A_C$ may be defined at the LHC as
\be
A_{C} = \frac{N(|y_t|>|y_{\bar t}|) - N(|y_t|<|y_{\bar t}|)}{N(|y_t|>|y_{\bar t}|) + N(|y_t|<|y_{\bar t}|)}.
\label{eq:def_afbt}
\ee
Measurements of $A_C$ at the LHC have been published by the CMS and ATLAS collaborations based on 
data sets with $4.7~{\rm fb}^{-1}$ of integrated luminosity~\cite{CMS-PAS-TOP-11-030,ATLAS-CONF-2012-057}.
The ATLAS central value is an order of magnitude larger than the CMS value, but they 
agree within the large uncertainties in both experiments, and they are consistent with the SM prediction.  

At the Tevatron, $t\bar t$ production is driven by the quark-antiquark initial state parton densities, whereas at the 
LHC, it is dominated by the gluon-gluon initial state which provides no asymmetry.   The overall asymmetry 
$A_C$ is therefore 
expected to be diluted substantially at the LHC.   An approximate estimate for the LHC asymmetry may be obtained by 
an extrapolation from the Tevatron result:   
\bea
A_C &\approx& \frac{\sigma(q\bar{q}\to t\bar{t})}{\sigma(gg\to t\bar{t})+\sigma(q\bar{q}\to 
t\bar{t})}\times A_{FB}^{t}(q\bar{q}\to t\bar{t}) \times \tilde\epsilon. 
\label{eq:ac_approx}
\eea
The first term represents the fraction of the top-quark pair production cross section 
induced by the $q\bar{q}$ initial state which is about $17~\%$ in the SM at the LHC 
at 7~TeV.  The second term is the asymmetry induced by the $q\bar{q}$ initial state. 
Given that about $88\%$ of the $t\bar t$ production cross section in the SM 
comes from the $q\bar q$ initial state at the Tevatron,  $A_{FB}^{t}(q\bar{q}\to t\bar
{t})$ can be extracted from the top quark forward-backward asymmetry observed 
at the Tevatron; we use $A_{FB}^{t}(q\bar{q}\to t\bar{t})\approx  A_{FB}^t /88\%$, where 
$A_{FB}^t$ is the measured top quark asymmetry. 
The last term $\tilde\epsilon$ in Eq.~(\ref{eq:ac_approx}) represents the probability 
of correct identification of  the forward direction, 
namely how frequently the forward direction represents the direction of the initial 
state quark.  This probability is evaluated in Berger{\em et al.}~\cite{Berger:2012tj} for 
both the Tevatron and the LHC.  

Combining all terms,  we expect  that  $A_C\simeq 0.17\times A_{FB}^{t}/88\% \times 
54\% \simeq 0.1 A_{FB}^t$, where $A_{FB}^t$ is the value measured at the Tevatron.
With $A_{FB}^t  \sim 20 \%$, an extrapolation from the Tevatron provides 
a model independent estimate for the LHC of 
$A_C^t  \simeq 0.02$, in reasonable agreement with the central value of the ATLAS 
measurement but in excess of the central value of the CMS measurement.   
Setting aside for the moment the still large uncertainties of the LHC data, the 
agreement of the ATLAS measurement with our extrapolation lends credence to 
the suggestion that new physics contributions are playing a role in the asymmetry 
measured at the Tevatron.  On the other hand, there is evident tension between 
the Tevatron asymmetry and the central value of the CMS measurement.  

\begin{figure}[!htb]
\includegraphics[scale=0.35]{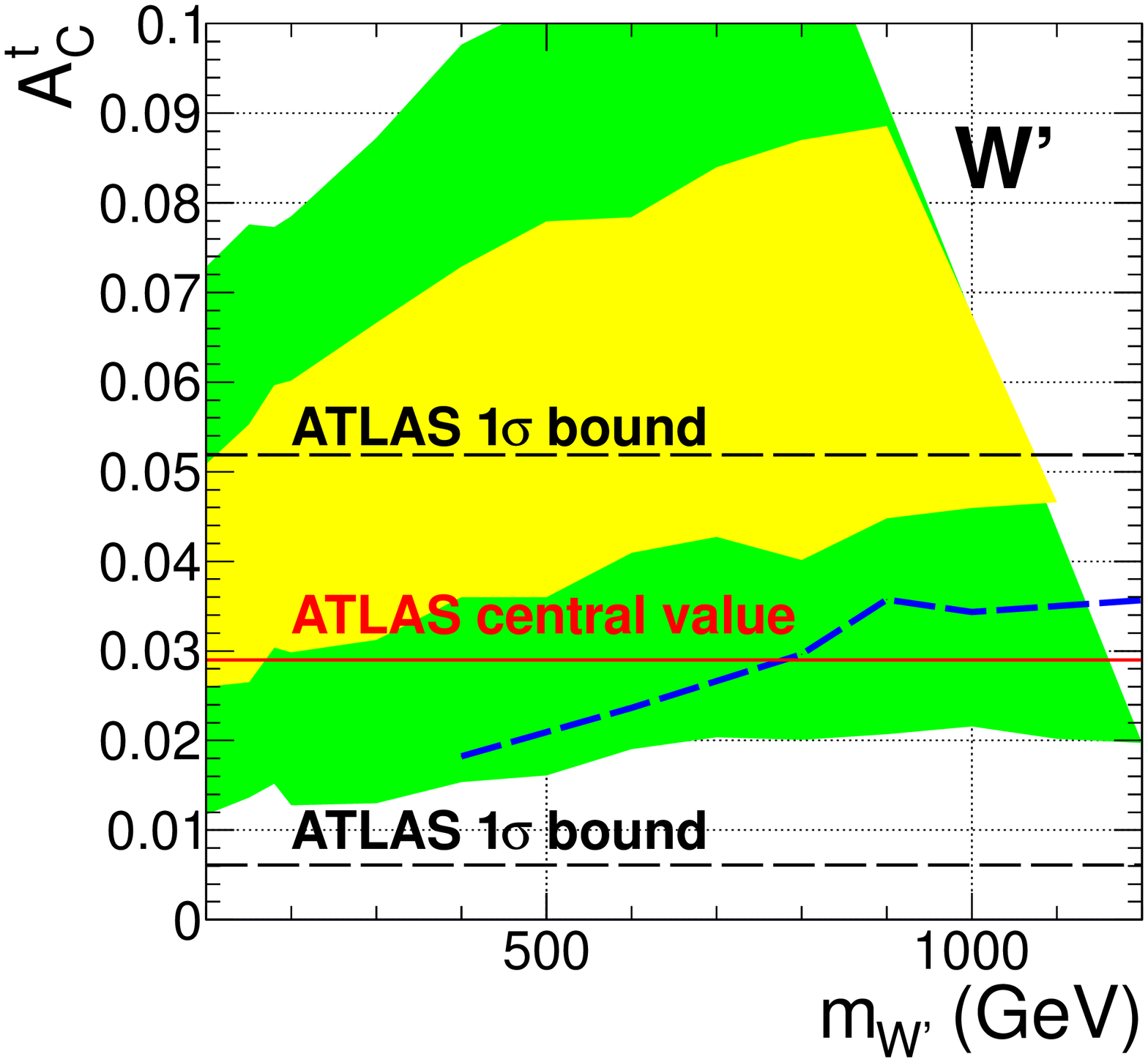}
\includegraphics[scale=0.35]{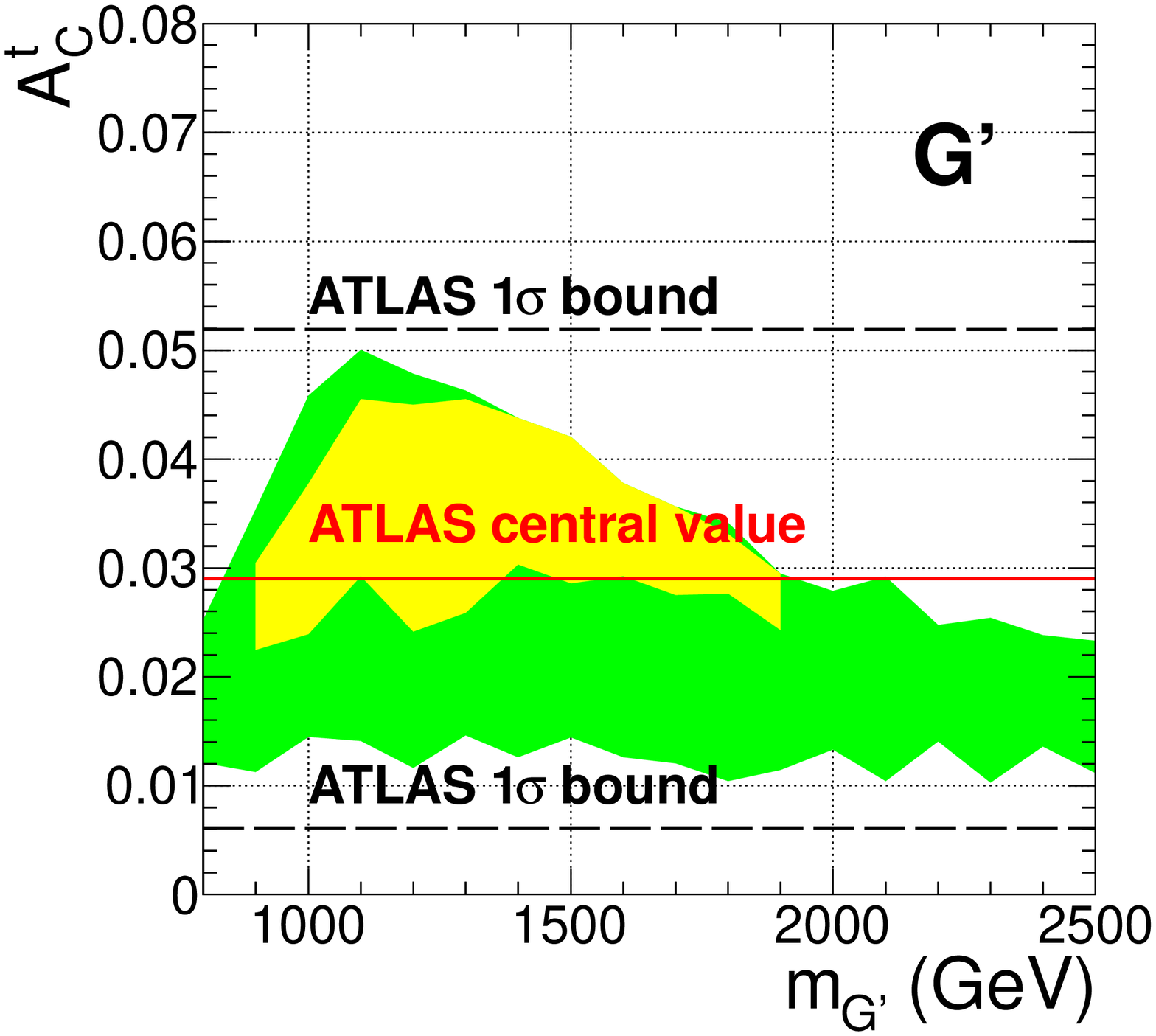}
\caption{The predicted top quark charge asymmetry, $A_C^t$, at the LHC at 7~TeV from the 
$W^\prime$ (left) and axigluon (right) models, compared with the ATLAS results. 
The yellow and green regions are for the couplings that 
fit the Tevatron $t\bar t$ total cross section and  $A_{FB}^t$ within $1\sigma$ and 
$2\sigma $, respectively.  The central value measured by ATLAS is marked with the 
red horizontal line, and the two black dashed lines 
show the $1 \sigma$ uncertainty of the measurement.  The blue dashed line on the 
$W^\prime$ figure shows the bound obtained from the analysis of 
top-pair-plus-one-jet events.  The region above this dashed line is disfavored.}
\label{fig:lhcActop}
\end{figure}

The extrapolation from the Tevatron is admittedly rough as it ignores possibly subtle 
energy-dependent effects and cancellations between SM and new physics contributions.  
Turning next to the explicit new physics models discussed in the previous section, we use 
the allowed parameters for the flavor-changing 
$W^\prime$ and axigluon models shown in Fig.~\ref{fig:tevfit} and calculate $A_C$  
at the LHC.   The results are shown in Fig.~\ref{fig:lhcActop},  along with a 
comparison to the ATLAS data.   To obtain the ATLAS predictions we 
use $A_C = 0.006$ for the SM prediction, as done by ATLAS.   For the CMS comparison, 
we use the SM value $A_C = 0.0115$ adopted by CMS.   The CMS comparison may be found 
in Berger {\em et al.}~\cite{Berger:2012tj}.  

Most of the values of $A_C$ predicted in the $W'$ model are larger than the ATLAS 
central value, but they are within the $1\sigma$ uncertainty band.   For the 
axigluon model, all of the predictions of $A_C$  agree with the ATLAS result within 
the $1\sigma$ level.  In the axigluon model $A_C$ does not simply increase with 
the axigluon coupling to SM particles.  For $m_{G'}=1500~{\rm GeV}$, $A_C$ 
reaches its maximum at about $4.2\%$, with coupling $f_{G'} =2.7$.  Therefore, 
the upper boundary of the yellow region (couplings that fit 
Tevatron data within $1\sigma$) overlaps the green region (couplings that fit 
Tevatron data within $2\sigma$) for some $m_{G'}$. 
The $G^\prime$ model predicts smaller values of $A_C$ than the $W^\prime$ model 
because there is a 
change of the sign of the {\it s-}channel propagator. When the invariant mass of 
the $t\bar{t}$ system is larger than the mass of the axigluon, the contribution to 
$A_C$ from the interference term is negative.   When    
comparing with the CMS data, we find in Berger {\em et al.}~\cite{Berger:2012tj} that the 
predicted values of $A_C$ are outside of the $1\sigma$ band.  
Unless the central value increases in updated measurements, the CMS data 
disagree with the simplest new physics models based on $W'$ or axigluon contributions. 

\section{The relationship of $A_{FB}^t$ and $A_{FB}^\ell$}
\label{sec:landtasy}

The top quark is the only quark that decays quickly, before hadronization takes place, 
and its polarization determines the kinematic distribution of its final state decay particles.  
Therefore, it should be possible to understand the relationship of $A_{FB}^t$ and 
$A_{FB}^\ell$based on the kinematics of the 
charged lepton in the decay of a top quark with different polarization states.  

The charged lepton in top quark decay is a powerful analyzer of the polarization of 
the top quark~\cite{Mahlon:2010gw}. Owing to the $V-A$ structure of the charged 
current in the SM, the angular distribution of a charged lepton $\ell^+$ from top quark 
decay ($t\to W^+ (\to \ell^+ \nu) b$) in the top quark rest frame  is 
\be
\frac{1}{\Gamma}\frac{d\Gamma}{d\cos\theta_{\rm hel}}=\frac{1+\lambda_t\cos\theta_{\rm hel}}{2},
\label{eq:spin}
\ee
where $\lambda_t$ denotes the top quark helicity, and $\theta_{\rm hel}$  is the angle of $\ell^+$ 
with respect to the direction of motion of the top quark in the overall center-of-mass system of the 
$t \bar{t}$ production process.  
We use the helicity basis in our calculations; $\lambda_t=+$ denotes  
a right-handed top quark ($t_R$),  and $\lambda_t=-$ a left-handed top quark ($t_L$). 
Once the top quark is boosted along its spin direction, the angular distribution of the 
charged lepton relative to the 
direction of motion of the top quark deviates from $(1\pm \cos\theta)$, and it becomes 
sensitive to the energy of the top quark $E_t$ (or equivalently its velocity $\beta$).  
We derive    
\begin{equation}
\frac{d\Gamma}{\Gamma d\cos\theta_{t\ell}}=\frac{1-\beta\cos\theta_{t\ell}+\lambda_t
\left(\cos\theta_{t\ell}-\beta\right)}{2\gamma^2\left(1-\beta\cos\theta_{t\ell}\right)^3}, 
\label{eq:lep_follow_top}
\end{equation}
where $\beta=\sqrt{1-m_t^2/E_t^2}$, $\gamma=E_t/m_t$, and $\theta_{t\ell}$ is the 
angle between the charged lepton and the direction of motion of its parent top quark.

To obtain the forward-backward asymmetry in the laboratory frame, we must rotate the angular distribution 
in Eq.~\ref{eq:lep_follow_top} from the top quark direction of motion to the laboratory coordinate axes.  
We use a function $R_F^{\ell,~\lambda_t}(\beta, y_t)$ to represent the probability that a lepton 
with positive charge lands in the forward region when it originates from a top quark with velocity 
$\beta$, rapidity $y_t$, and polarization $\lambda_t$.    Formally, 
\be
R_F^{\ell,~\lambda_t}(\beta, y_t)=\frac{N_F^\ell}{N_F^\ell+N_B^\ell}.  
\label{eq:defR}
\ee
where $N_F^\ell$ ($N_B^\ell$) denotes the number of leptons $\ell$ in the 
forward (backward) region in the laboratory.   Moreover,
\be 
A_{FB}^{\ell,~\lambda_t}(\beta, y_t) = 2 R_F^{\ell,~\lambda_t}(\beta, y_t) - 1 .
\label{eq:defA}
\ee
It is noteworthy that an explicit analytic expression can be obtained in closed form for 
$R_F^{\ell,~\lambda_t}(\beta, y_t)$ in the laboratory frame.   The derivation is somewhat lengthy, 
and it is presented in the Appendix of Berger {\em et al.}~\cite{Berger:2012tj}.   

The functions $R_F^{\ell,~\lambda_t}(\beta, y_t)$ in Eq.~\ref{eq:defR} and 
$A_F^{\ell,~\lambda_t}(\beta, y_t)$ in Eq.~\ref{eq:defA} 
are functions of the top quark momentum.   To obtain the numbers of leptons 
in the forward and backward regions, we must convolve $R_F^{\ell,~\lambda_t}
(\beta, y_t)$ with the top quark momentum spectrum, i.e. 
\begin{eqnarray}
\frac{N^\ell_F}{N^\ell_F+N^\ell_B}&=&\frac{1}{\sigma}\sum_{\lambda=+,-}
\int R_F^{\lambda}\left(\beta,y_t\right)
\frac{d^2\sigma|_{\lambda_t=\lambda}}{d\beta dy_t}d\beta \wedge dy_t,\\
\frac{N^\ell_B}{N^\ell_F+N^\ell_B}&=&\frac{1}{\sigma}\sum_{\lambda=+,-}
\int\left[1-R_F^{\lambda}\left(\beta,y_t\right)\right]\frac{d^2\sigma|_{\lambda_t=\lambda}}
{d\beta dy_t}d\beta \wedge dy_t,\\
A_{FB}^\ell&=&\frac{1}{\sigma}\sum_{\lambda=+,-}\int \left[2R_F^{\lambda}
\left(\beta,y_t\right)-1\right]\frac{d^2\sigma|_{\lambda_t=\lambda}}{d\beta dy_t}
d\beta \wedge dy_t
\label{eq:correlation}
\end{eqnarray}
where $\displaystyle{\frac{d^2\sigma|_{\lambda_t=\lambda}}{d\beta dy_t}}$ labels the 
differential $t\bar{t}$ production cross section for a top quark with specific kinematics
($\beta$, $y_t$, $\lambda_t$), and $\sigma$ stands for the $t\bar{t}$ total
production cross section. 

The observed positive top-quark asymmetry $A_{FB}^t$ indicates that more top 
quarks are produced  in the forward region than in the backward region of rapidity.   
Both $t_R$ and $t_L$ can generate a positive lepton asymmetry $A_{FB}^\ell$ from 
a positive $A_{FB}^t$ . However, a $t_L$ would need a large boost along the proton beam 
line (i.e. in the large forward rapidity region) to overcome the fact that most of the 
charged leptons from its decay move against it in its rest frame. A right-handed top 
quark $t_R$ can yield a positive $A_{FB}^\ell$ even for top quarks near the $t\bar{t}$ 
threshold region.  Therefore, the large positive top quark and lepton asymmetries 
$A_{FB}^t$ and $A_{FB}^\ell$ observed by the D0 collaboration indicate that
the top quark polarization and the kinematics of the top quarks, $y_t$ and $E_t$, may 
be playing a non-trivial role.   

\begin{figure}[!htb]
\includegraphics[scale=0.35]{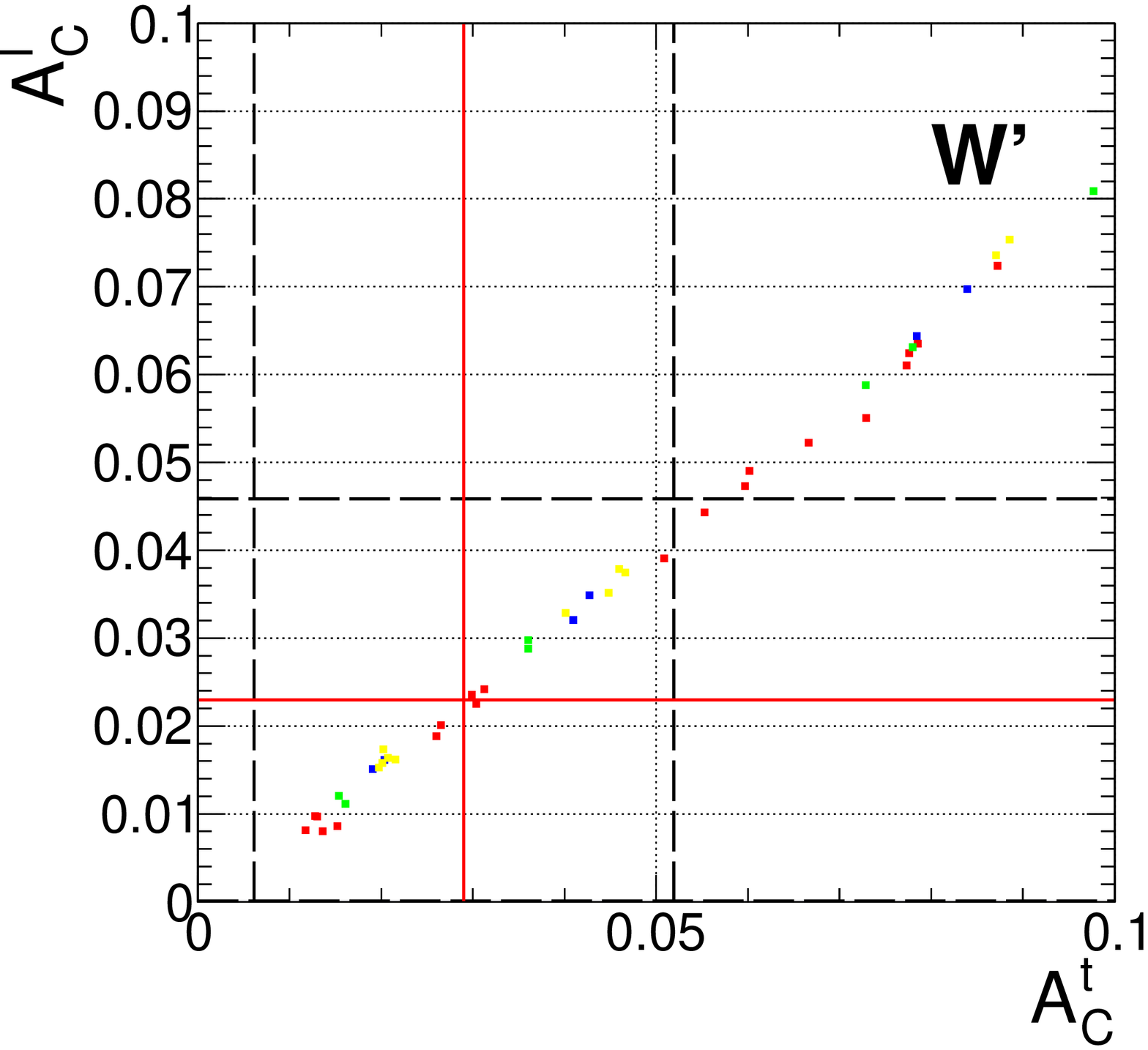}
\includegraphics[scale=0.35]{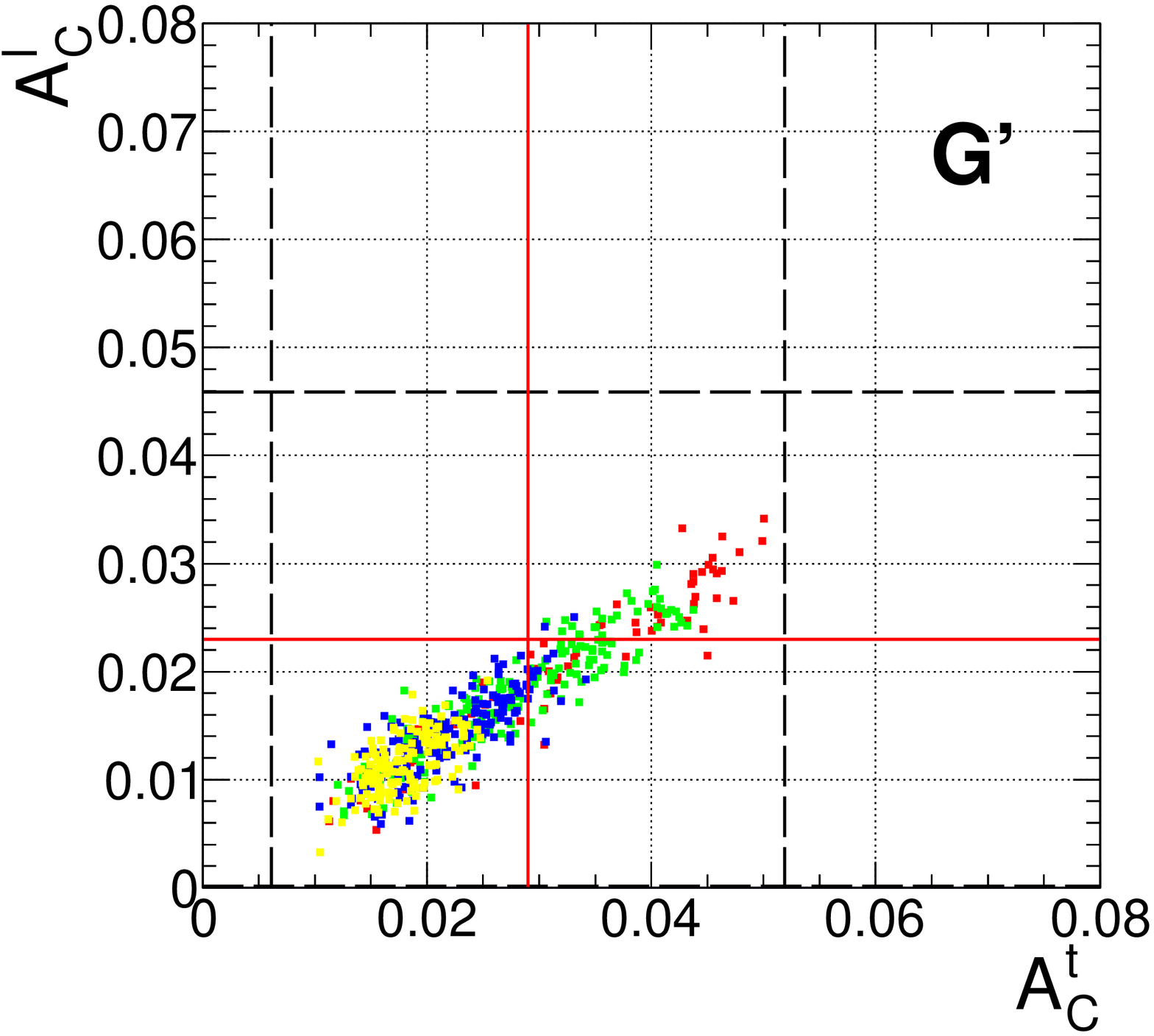}
\caption{The correlation between 
$A_C^t$ and $A_C^\ell$ at the LHC for the $W^\prime$ (left) and axigluon models 
(right).  The vertical (horizontal) red line and the two black dashed lines show the central 
value of $A_C^t$ ($A_C^\ell$) and the $1\sigma$ uncertainty bands measured by ATLAS 
at the LHC.  The green (red) dots are obtained from the parameters that fit the Tevatron 
$t\bar t$ cross section and $A_{FB}^t$ within $1\sigma$ ($2\sigma$).}
\label{fig:correlation}
\end{figure}

The correlation between the charged lepton asymmetry and the top quark 
asymmetry is significantly different for different polarization states of the top 
quark, and it may therefore shed light on the nature of the physics that 
causes the forward-backward asymmetries at the Tevatron.   We choose the 
$W^\prime$ 
and axigluon models as two reference models to examine the correlation at 
the Tevatron and the LHC.   

The axigluon and  $W^\prime$ models admit good fits to $A_{FB}^t$ at the Tevatron, 
but they provide distinct predictions for the polarization and kinematics of the final 
state top quark.  The $W'$ model produces dominantly $t_R$ while the axigluon 
model generates an equal number of $t_R$ and $t_L$ with more energetic top 
quarks since the quarks come from the decay of a heavy axigluon. In Fig.~\ref{fig:correlation}, 
we show the results of our calculation of the charged lepton asymmetry at the LHC 
using the parameters 
determined in our $1\sigma$ fits to the $t\bar t$ total cross section and the most recent 
CDF data on $A_{FB}^t$.   Figure~\ref{fig:correlation} shows 
charged lepton asymmetry for the LHC together with the top quark charge asymmetry 
$A_C^t$.    The results for the Tevatron are shown in Berger {\em et al.}~\cite{Berger:2012tj}.  
There are vertical red lines in Fig.~\ref{fig:correlation} at $A_C^t \sim 0.03$ to show the 
central values of the asymmetries measured by ATLAS, and two black dashed 
lines show the extent of the quoted experimental 
$1\sigma$ uncertainty bands.   The horizontal red line shows the central 
value of $A_C^\ell$ measured by ATLAS, and the horizontal black dashed lines 
show the $1\sigma$ uncertainty values.    

The predicted charged lepton asymmetries stretch out over a range of values depending 
on the values of the axigluon or $W'$ masses used in the fits to the Tevatron data.   At the 
LHC, there are parameters in both models (obtained from the Tevatron fits) that can reproduce 
the values of $A_C^t$ and $A_C^\ell$ measured by ATLAS, shown by the fact 
that the intersection of the vertical and horizontal red lines passes through the scattering of dots. 
On the other hand, there is a wide range of dots in the $W'$ model that are above the central 
values of $A_C^t$ and $A_C^\ell$, and out of the $1\sigma$ uncertainty band.  In the 
axigluon model, all the values of $A_C^t$ and $A_C^\ell$ are consistent with ATLAS 
measurements within the $1\sigma$ bands.   The LHC and Tevatron data 
together could reduce the allowed parameter spaces of the two models.  

The size of the top quark asymmetry, in excess of SM expectations,  is one indication 
that new physics may be playing a role. The charged lepton asymmetry provides a second 
and independent indication of the presence of new physics since it points to the possibility 
that more right- than left-handed top quarks are being produced.   It is important to confirm the charged 
lepton asymmetry.   This goal could be realized with an analysis of the full data set in D0.  It would 
be valuable also to have a measurement of the charged lepton asymmetry from the CDF collaboration.
 
\section*{Acknowledgments}
The work reported here was done in collaboration with Qing-Hong Cao, Chuan-Ren Chen, 
and Hao Zhang.  The High Energy Physics Division at Argonne is supported by the U.S. DOE 
under Grant No.~DE-AC02-06CH11357. 

I am pleased to commend Marc Besancon for his very professional and self-effacing organization of 
the excellent scientific program of this conference on Beyond the Standard Model of Particle Physics in 
Quy Nhon, Vietnam in July 2012.   I am glad to have had this opportunity to make his acquaintance.  

This conference was the most recent in of the ``Rencontres du Vietnam'' series established by 
Jean Tran Thanh Van.  Van has contributed much over the years to fostering international scientific 
collaboration, and to facilitating interactions among experimenters and theorists, through his 
establishment of the ongoing ``Rencontres de Moriond" series begun in 1966 and the ``Rencontres 
de Blois" series started in 1989.  During this visit to Vietnam, conference participants had the opportunity to 
learn first-hand of the ambitious International Center for Interdisciplinary Science and Education that 
he is creating in Quy Nhon, Vietnam and to visit the sea-side construction site.   Once operational, the 
intent is that the Center will host high-level national, regional, and international meetings in basic and 
applied science, medicine, the humanities, and social sciences.

\end{document}